\documentclass{article}
\usepackage{spconf,amsmath,graphicx}

\title{SEQUENTIAL MONTE CARLO GRAPH CONVOLUTIONAL NETWORK \\
FOR DYNAMIC BRAIN CONNECTIVITY}
\name{Fengfan Zhao, Ercan Engin Kuruoglu}
\address{Tsinghua-Berkeley Shenzhen Institute, Shenzhen International Graduate School, Tsinghua University}
\begin{document}
%
\maketitle
\begin{abstract}
An increasingly important brain function analysis modality is functional connectivity analysis which regards connections as statistical codependency between the signals of different brain regions. Graph-based analysis of brain connectivity provides a new way of exploring the association between brain functional deficits and the structural disruption related to brain disorders, but the current implementations have limited capability due to the assumptions of noise-free data and stationary graph topology. We propose a new methodology based on the particle filtering algorithm, with proven success in tracking problems, which estimates the hidden states of a dynamic graph with only partial and noisy observations, without the assumptions of stationarity on connectivity. We enrich the particle filtering state equation with a graph Neural Network called Sequential Monte Carlo Graph Convolutional Network (SMC-GCN), which due to the nonlinear regression capability, can limit spurious connections in the graph. Experiment studies demonstrate that SMC-GCN achieves the superior performance of several methods in brain disorder classification.
\end{abstract}
\section{Introduction}
\label{sec:intro}

Functional dynamics include changes in the strength of connections between regions, and also the number of connections linked to regions. Various works \cite{faisan2007hidden, hutchinson2009modeling, janoos2011spatio} independently devised different types of state-space models to explore the dynamic characteristics of functional activation and applied them to event-related fMRI data analysis. Brain connectivity can be quantified by encoding neighbourhood relations into a connectivity matrix, the rows and columns of which correspond to different brain regions of interest (ROI). This representation lends itself to be mapped to a graphical model which provides means to quantify different topological aspects of the connectome. Recent years have witnessed an exponential growth of studies on the applications of Graph Convolutional Networks (GCNs) in neuroscience, in particular in finding common patterns or biomarkers \cite{BrainGNN}. 
 
The GCN-based methods on fMRI data can be categorized into two subgroups depending on the definition of nodes in the graph, i.e., population graph-based models and brain region graph-based models \cite{GNNinNN}. These two graph models correspond to two separate tasks for fMRI analysis, the node classification tasks and graph classification tasks. In population graph-based models, the nodes in the graph denote the subjects and the edges represent the similarity between subjects.  PopulationGCN \cite{population_gcn}  involves representing populations as a sparse graph in which nodes are associated with imaging features and edge weights are constructed from phenotype information (e.g., age, gene, and sex of the subjects). 
One of the methods, BrainGNN \cite{BrainGNN}, conducts an interpretable GCN model on a brain region graph to understand which brain regions are related to a specific neurological disorder. 

Both of these graph-based models only used static functional connectivity information, where pairwise correlations between regions are calculated using the entire duration of the fMRI scan. This neglects the fact that brain connectivity is dynamic, with functional connectivity fluctuating over time. Dynamic connectivity presents time-varying region connections as opposed to static connectivity, which is becoming the frontier in fMRI data discovery \cite{PRETI201741}.

A dynamic system estimation problem requires a dynamic model estimation method. Sequential Monte Carlo (SMC) provides a solution to dynamic system tracking problems by estimating the hidden states of a dynamic system with only partial and noisy observations and had important success in various applications \cite{doucet2001sequential, costagli2007image}.
A specific SMC methodology called Particle Filter (PF) provides an extension to Kalman filtering to nonlinear systems and possibly non-Gaussian time-series and noise by employing a Monte Carlo sampling approach to iteratively track the posteriors of hidden variables of a system of interest using observations under noise. PF was applied to simulated recordings of electrical and neurovascular mediated hemodynamic activity, and the advantages of a unified framework were shown \cite{Croce_2017}. Ancherbak et al. implement PF application on gene network dynamics \cite{ancherbak2015time} and Ambrosi et al. \cite{Ambrosi2021} apply PF and vector autoregressive model in brain connectivity for dynamic modelling. Despite encouraging results obtained with these methods, the graph topological information is not utilised beyond a linear vector regression model for hidden variables.

In this work, we provide an extension to the nonlinear state evolution model while exploiting graph topological information by utilising a Graph Convolutional Network for uncovering connectomes from fMRI images of the brain. Our method, namely Sequential Monte Carlo Graph Convolutional Network (SMC-GCN), applies the particle filter to graphs with GCN backbone providing dynamic connectivity robustness and avoiding spurious links. The experiment result demonstrates that SMC-GCN achieves superior performance compared to other dynamic connectivity methods for brain disorder classification.

\section{Methodology}

\begin{figure*}[ht]
    \centering
    \includegraphics[width=1.78\columnwidth]{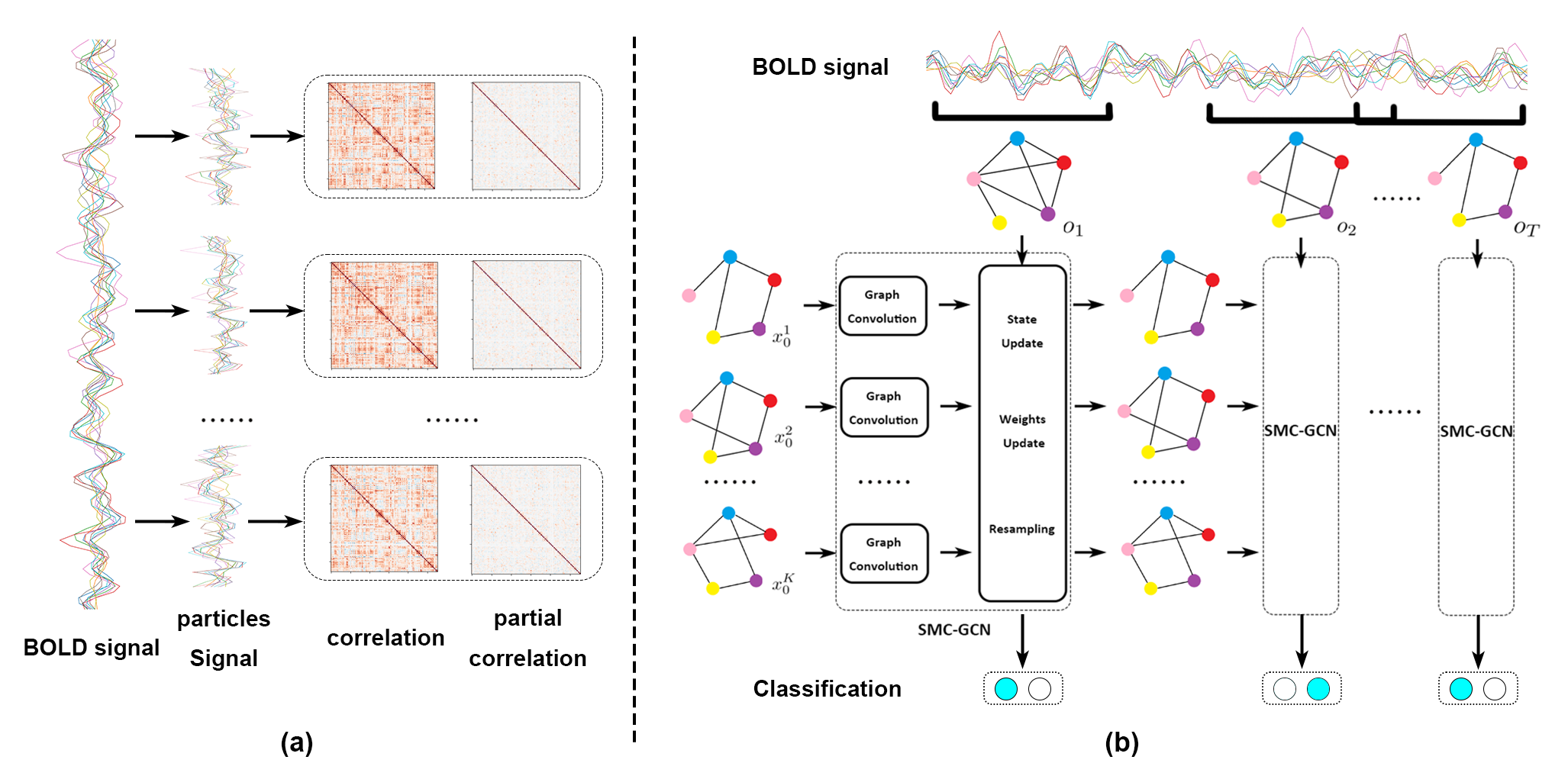}
    \caption{\small Overall Architecture of SMC-GCN for Dynamic Brain Connectivity. (a) The fMRI time series is split into $T$ time series by sliding window approach \cite{mokhtari2019sliding} for graph construction. After that, $K$ graph particles $\{ \boldsymbol X_t^k\}_{t=0:T}^{k=1:K}$ with initial particle weights $w^k = 1/K$ are constructed. (b) $K$ graph particles are fed into SMC-GCN simultaneously for state transition acquiring priors for the next timestamp $\boldsymbol X_{t+1}^k$, following the state observation by $\boldsymbol O_{t+1}^k$ and resampling over time. Each timestamp outputs the readout classification result.}
    \label{model}
\end{figure*}
\subsection{Particle Filtering}

Given the system model:
\begin{equation}
\begin{aligned}
    x_{0} &\sim p_0(x) \\
    x_{t} &= g_t(x_{t-1},v_t), \text{ for } t \geq 1, \\
    o_t &= h_t(x_t,w_t), \text{ for } t \geq 1
\end{aligned}
\end{equation} 
where $v_t$ and $w_t$ are the system transition and observation noise, $g_t$ and $h_t$ are functions of system. Consider a Markov process with the state transition probability given by $p(x_t|x_{t-1})$ where $x_t$ is the state at the timestamp $t$. For every transition, we receive an observation $o_t$ described by the observation probability of $p(o_t|x_t)$. After a sequential of observations $o_{t=1:T}= \{o_1,\dots, o_T \}$, particle filtering estimates the posterior probability $p(x_T |o_{t=1:T} )$ of the state $x_T$ conditioned on the sequence of observations. Particles are sequentially updated with three steps: First is the state transition step with the previous posterior state and transition model,
\begin{equation}
p(x_t | o_{1:t-1}) = \int p(x_t|x_{t-1})p(x_{t-1}|o_{1:t-1}) d x_{t-1}
\end{equation}
Then is the state update step with the current observation and transition state, 
\begin{equation} 
p(x_{t}|o_{1:t}) = \frac {p(o_{t}|x_{t})p(x_{t}|o_{1:t-1})} {p(o_{t}|o_{1:t-1})}
\end{equation}
as well as the weight update following the proposal density $q(\cdot)$ rather than the posterior observation $p(\cdot)$. 
\begin{equation}
    w_t^k \propto w_{t-1}^k \frac{p(o_t|x_t^k)p(x_t^k|x_{t-1}^k)}{q(x_t^k|x_{t-1}^k,o_t)}
\end{equation}
The last is the resampling step. To avoid the degeneracy problem, particles need to resample $K$ from the discrete distribution of $\langle w^k_{t+1} \rangle_{k=1:K}$ and reset weights to uniform distribution $1/K$. 

\subsection{Sequential Monte Carlo Graph Convolutional Network}

Most of the existing work utilizes linear state updates for uniform or multivariate variables. In this work, we apply particle filtering on graphs where the observables are the node features while the state variable is the graph structure.
Unlike classical particle filtering usage, we use a nonlinear model for state changes for which we utilize a Graph Convolutional Network.
In this setup, since the state variables are graphs, the particles become graphs themselves to which weights are assigned.
Rather than vectors, graphs are sampled forming {\em graph particles}.
The overall architecture is shown in Fig. \ref{model}.

\subsubsection{Graph Construction}
For initialization, graph particles are sampled from the ROI time series with the length $\Gamma$ timestamps. Each graph particle $\boldsymbol X_t^k = \{A_t^k, F_t^k\}$ is constructed by the node of ROIs defined by a specific atlas. The adjacency matrix $A$ is computed by partial correlations between ROIs for the sparse weighted connection between nodes and the feature embeddings $F$ corresponding to each ROIs are the correlation computation between ROIs for further transition and observation steps.

To construct a sequence of dynamic graph observations, we utilize the popular sliding window approach \cite{mokhtari2019sliding}. Given a window length $\Gamma$, and stride $S$, the $T = \lfloor T_{\text{max}} - \Gamma / S\rfloor $ windows are constructed. Each observation $\boldsymbol O_t = \{A_t, F_t\}$ has the same construction with states.

\subsubsection{State Transition}

We make the GCN model for state transition, motivated by the Particle Filtering Network \cite{karkus2018particle} replacing the observation model using CNN for visual localization. For the state transition step, each graph particle at timestamp $t-1$ will feed into the GCN backbone as the next state estimation where the system noise is simulated by different graph particles. It is the prior calculation for the next timestamp $t$ state, the same as the GCN learning and prediction. The state transition equation is shown as follows:

\begin{equation}
    \boldsymbol X_t^k | \boldsymbol O_{1:t-1}^k = \text{GCN}(\boldsymbol X_{t-1}^k|\boldsymbol O_{1:t-1}^k;\theta)
\end{equation}
where $\theta$ is the learnable parameter in GCN with shared parameters during $T$ timestamp transition and observation. Any specific convolution operation including Chebyshev polynomials \cite{defferrard2016convolutional} or simple GCN convolution \cite{chen2020simple} can be applied to the GCN model.

\subsubsection{State Update}
The state update step is the \textit{posterior} calculation that the input graph particle at the timestamp $t$ $\boldsymbol X_t^k$ is updated by the current observation $\boldsymbol O_t$. Firstly, the adjacency matrices are aggregated by the top-K largest observation adjacency $A_t^k = A_t^k(\boldsymbol X) + \text{TopK}(A_t(\boldsymbol O))$ and the feature matrices are modified to $F_t^k =F_t^k(\boldsymbol X) \circ F_t^k(\boldsymbol O) $ with the Hadamard product. The state update aggregation is shown as follows:
\begin{equation}
    \boldsymbol X_{t}|\boldsymbol O_{1:t} = \text{AGGREGATE}(\boldsymbol O_t,\boldsymbol X_t|\boldsymbol O_{1:t-1})
\end{equation}

In the absence of intermediate probability density for graph observation, we have used discriminative function $f(\boldsymbol O_t, \boldsymbol X_t^k)$ to up/down-weight the state. This function discriminates the differences in node features between observations and predicted states. First, it aggregates the neighbour features using $K$ nearest neighbours and then makes the comparison with the predicted state.
\begin{equation}
f(\boldsymbol O_t, \boldsymbol X_t^k) = \exp \left\{ \sum_{v \in \mathcal{V}}\log p(\boldsymbol O_{t,v}|\boldsymbol X_t^k)\right\}
\end{equation}
$\log p(\boldsymbol O_{t,v}|\boldsymbol X_t^k)$ denote the log-likelihood of the measurement at node $v$ and time $t$. Since it is discriminatively trained to optimize the end task, it performs the same function as $p(o|x)$ in particle filtering. 
\begin{equation}
w_{t}^k = \frac{w_{t-1}^k f(\boldsymbol O_t, \boldsymbol X_t^k)}{\sum_k w_{t-1}^k f(\boldsymbol O_t, \boldsymbol X_t^k)}
\end{equation}
The $\sum_k w_{t-1}^k f(\boldsymbol O_t,\boldsymbol X_t^k)$ is the normalization term.

\subsubsection{Soft Resampling}

For the particle filtering resampling step, the weights are adjusted to the uniform distribution. Since this is a non-differentiable operation, we adopt the differentiable soft-resampling strategy \cite{karkus2018particle}. Specifically, we sample new particles from the convex combination of particle weights and uniform distribution, 
\begin{equation}
    {w'}_t^k=\frac{p_t(k)}{q_t(k)}=\frac{w_t^k}{\alpha w_t^k +(1-\alpha)1/K}
\end{equation}
where $\alpha \in [0, 1]$ is a tunable parameter. 
This differentiable approximation provides non-zero gradients for the full particle chain with a trade-off between the desired sampling distribution ($\alpha = 1$) and the uniform sampling distribution ($\alpha = 0$).

\begin{table*}[ht]
  \label{tab:result}
  \centering
  \caption{Comparison of Cross-Validation Results on ABIDE Dataset} 
  \begin{tabular}{l|c c c c}
    \hline
    Method & Accuracy & Sensitivity & Specificity & AUC\\
    \hline
    SVM & 61.87($\pm$0.36) & 65.74($\pm$0.24) & 69.54($\pm$0.26) & 67.74($\pm$0.71) \\
    RF & 60.58($\pm$0.58) & 52.30($\pm$0.91) & \textbf{74.36($\pm$0.83)} & 65.31($\pm$0.24) \\
    GCN & 68.28($\pm$0.84) & 60.25($\pm$0.63) & 71.54($\pm$0.46) & 71.96($\pm$0.53) \\
    GAT & 66.24($\pm$0.73) & 63.21($\pm$4.76) & 69.45($\pm$0.78) & 69.87($\pm$0.47) \\
    BrainGNN & 70.66($\pm$0.56) & 67.30($\pm$0.99) & 71.98($\pm$1.44) & 72.71($\pm$0.87) \\
    \hline
    SMC-GCN & \textbf{73.64($\pm$0.98)} & \textbf{70.06($\pm$1.56)} & 73.48($\pm$1.28) & \textbf{75.29($\pm$1.01)} \\
    \hline
\end{tabular}
\end{table*}

\subsection{Graph Particle Classification}

Lastly, we seek a flattening operation to preserve information about the input graph in a fixed-size representation. Concretely, to summarize the output graph of the convolutional global pooling block, we use both mean and max pooling:
\begin{equation}
    z_t^k = \text{mean} (\boldsymbol H_t^k) \| \text{max} (\boldsymbol H_t^k)
\end{equation}
where $H$ is the hidden embeddings during each transition GCN, with mean and max operating elementwisely, and $\|$ denotes concatenation. To retain information on a graph in a vector, we concatenate both mean and max summarization for a more informative graph-level representation. The final prediction is submitted to an MLP to obtain the graph classification with particle weights as average results.
\begin{equation}
    \hat{y}_{t, c} = \sum_{k=1} ^{K} w^k \cdot \text{MLP}(z_t^k)
\end{equation}
Cross-entropy loss is used in the final classification for all timestamps $T$.

\section{Experiments}

\subsection{Dataset}
We apply our model to the large and challenging database for binary classification tasks. The Autism Brain Imaging Data Exchange (ABIDE) database \cite{ABIDE_preprocessed} aggregates data from different acquisition sites and openly shares functional MRI and phenotypic data of 1112 subjects from across 20 international sites.  To ensure a fair comparison with the previous works, we use the same preprocessing pipeline, the Configurable Pipeline for the Analysis of Connectomes (C-PAC), with band-pass filtering and no global signal regression. We take the 871 subjects that passed manual quality control checks from three expert human reviewers and further prune the data with less than 160 time steps to a final total of 578 samples. We use the Havard-Oxford (HO) atlas for ROI splitting. The resulting class distribution is 287 subjects with ASD to 291 healthy controls. 

\subsection{Experiment Setting}
We utilize a stratified 5-fold cross-validated training procedure with a data split of 60\% training, 20\% validation, and 20\% testing during hyperparameter tuning. The stratification occurs over both class (ASD and HC) and site (one of 17 scanning locations), in an attempt to minimize the effects of the different scanning parameters on our All training and testing are conducted on one NVIDIA RTX 3090 GPU. We use two layers of Chebyshev convolution with polynomials $K=3$. We train our model using the Adam optimizer with the learning rate of $0.01$ and the batch size is set to $12$. The sliding window size is $100$ and the stride is fixed for timestamp $T=20$. The number of particles is $K = 30$.

The 5-fold cross-validation results of our methods are presented in Table 1, with classification performance compared to previous results. The baseline models consist of machine learning methods including SVM, Random Forest (RF) and graph based approaches including GCN, Graph Attention Networks (GAT) \cite{veličković2018graph} and BrainGNN \cite{BrainGNN}. We achieve superior performance among baseline models.


\begin{figure}
    \centering
    \includegraphics[width=.8\columnwidth]{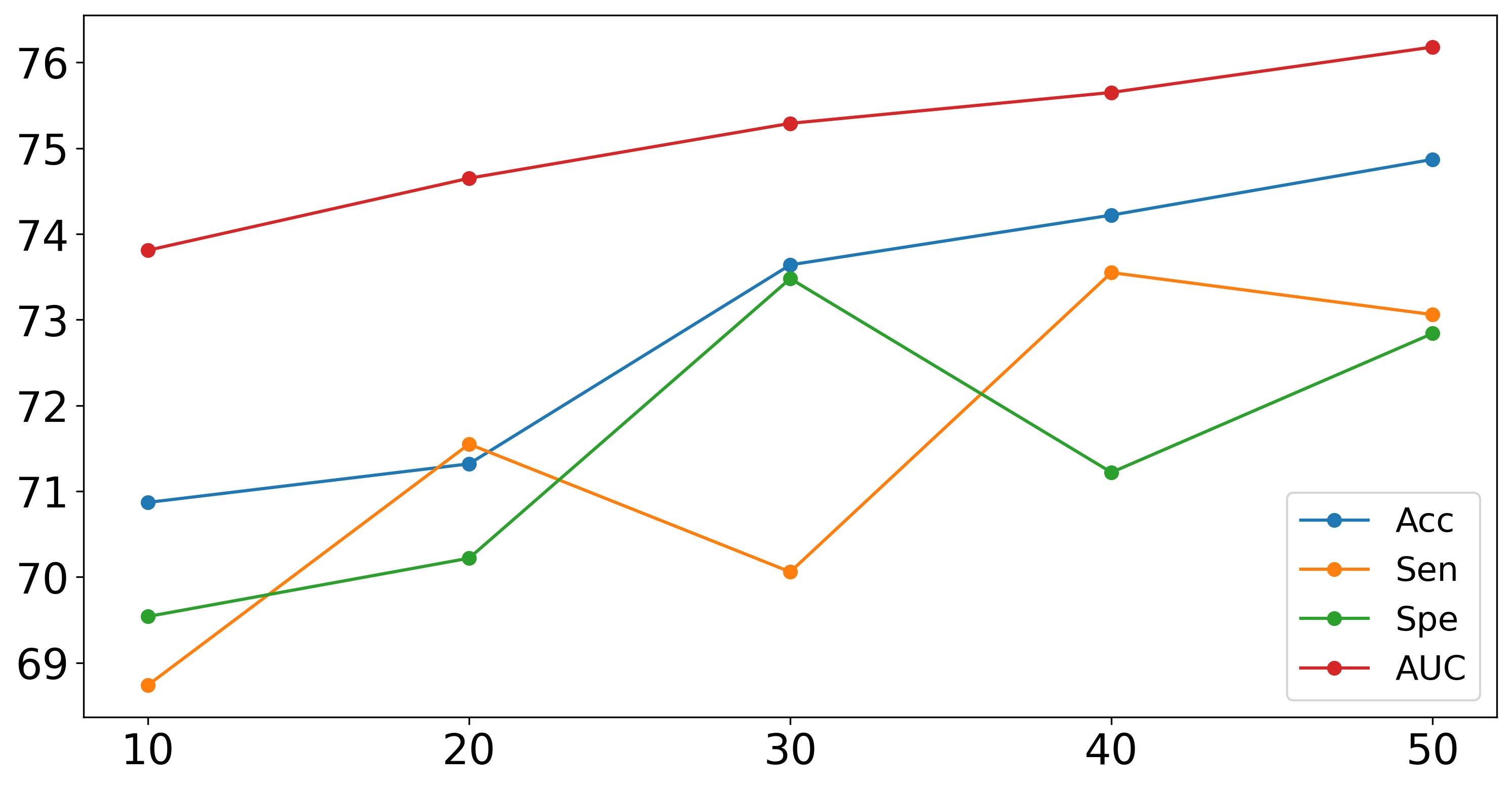}
    \caption{Number of Particles Performance}
    \label{fig:enter-label}
\end{figure}

\subsection{Ablation Study}
We mainly focus on the number of particles ablation. For traditional particle filtering algorithms, particles sampled from pdf perform better when the particle number explodes. So for graph particles, we sample 10 to 50 particles in experiments due to training graph size. The results are shown in Table 2, with smooth growth of the accuracy and AUC score, while the training time is doubled, compared to number = 10. Since no thresholds appear when the particle number grows, we will further investigate the efficiency with the growth of the particle number for sampling approximation.

\section{Conclusion}

In this work, we have extended the classical particle filtering to a broader interpretation where the particles could correspond to graphs or other data embedding structures.
We have further enriched the learning process with neural networks leading to the 
the Sequential Monte Carlo Graph Convolutional Network (SMC-GCN). This work has been motivated by the need to follow brain dynamics and has been used to model sequential connectivity
in brain disorders.
The method demonstrates superior performance compared to the baseline methods. 
Our future work will explore other applications of the SMC-GCN model with other dynamic graph topologies and its capacity for the growth of particles.

\pagebreak

\bibliographystyle{IEEEbib}
\bibliography{strings}

\end{document}